\documentclass{eas}
\usepackage{graphicx}
\usepackage{hyperref}

\hypersetup{
    bookmarks  = true,
    pdftitle   = {Magnetic Field Induced Radius Inflation of Low-Mass Stars},
    colorlinks = true,
    linkcolor  = blue,
    citecolor  = blue
}
    
\newcommand{\msun}{M_\odot}

\TitreGlobal{Binary 2013: Setting a new standard in the analysis of binary stars.}
\begin{document}

\title{Magnetic Field Induced Radius Inflation of Low-Mass Stars}
\runningtitle{Do Magnetic Fields Inflate Low-Mass Stars?}
\author{Gregory A. Feiden}
    \address{Dept.\ of Physics \& Astronomy, Uppsala University, Box 516, 
             SE-751 20 Uppsala, Sweden}
    \secondaddress{Dept.\ of Physics \& Astronomy, Dartmouth College,
                   6127 Wilder Laboratory, Hanover, NH 03755, USA.}
\author{Brian Chaboyer}\sameaddress{2}
\begin{abstract}
We present results obtained using the magnetic Dartmouth stellar
evolution code that address the possibility that magnetic
fields are inflating low-mass stars in detached eclipsing binaries. 
While it seems plausible that magnetic fields are inflating stars with
radiative cores, the level of inflation observed among fully convective
stars appears too large to be explained by magnetic fields. We provide
an alternative explanation, stellar metallicity, and propose observations
that can help further constrain stellar models.
\end{abstract}

\maketitle

\section{Introduction}
Observations of detached eclipsing binaries (DEBs) with at least one low-mass
component ($M < 0.8\msun$) have shown, quite convincingly, that stellar
evolution models do a poor job of predicting fundamental properties
of low-mass stars (e.g., Ribas \cite{Ribas2006}, Feiden \& Chaboyer \cite{FC12a}). 
Models routinely under predict radii and over predict effective temperatures 
by approximately 4\% and 3\%, respectively. It is often suggested that the 
interaction of stellar magnetic fields with thermal convection belies these 
noted discrepancies (e.g., Mullan \& MacDonald \cite{MM01}, Chabrier \etal\ 
\cite{Chabrier2007}). Support is lent to this suggestion
by the observation that magnetic activity indicators appear to correlate with
radius discrepancies (e.g., L\'{o}pez-Morales \cite{Lopez2007}, Feiden \& Chaboyer
\cite{FC12a}). We have recently 
initiated an effort to test this hypothesis using the Dartmouth stellar 
evolution code (Dotter \etal\ \cite{Dotter2007}, Feiden \& Chaboyer \cite{FC12b}). 
In these proceedings, we will highlight some
current results of this effort and discuss what we feel are key 
observations to identify the mechanism causing these discrepancies.

\section{Magnetic Stellar Evolution Models}
Effects of a magnetic field have been incorporated into the Dartmouth stellar
evolution code in a self-consistent manner (Feiden \& Chaboyer \cite{FC12b}).
At the moment, magneto-convection is included using two different prescriptions 
(Feiden \& Chaboyer \cite{FC12b}, \cite{FC13a}). We will refer to them as:
(1) {stabilization of convection}, and (2) inhibition of convective 
efficiency. Stabilization of convection refers to a self-consistent modification
to the Schwarzschild stability criterion that is brought about by assuming
the magnetic field is in thermal equilibrium with the surrounding plasma. It
is similar in effect to magneto-convection implemented by Mullan \& MacDonald
(\cite{MM01}). The second method, inhibition of convective efficiency, assumes
that the energy required to create a magnetic field is drained from kinetic
energy in convection (see Feiden \& Chaboyer \cite{FC13a} for details). 
Physically, it can be likened to reduced convective mixing length methods 
(e.g., Chabrier \etal\ \cite{Chabrier2007}).

\section{Results}
We will now highlight general conclusions drawn from applying the aforementioned 
magnetic stellar evolution models to a small number of well characterized DEBs. 
A full account of the results for each individual system can be found elsewhere
(Feiden \& Chaboyer \cite{FC12b,FC13a}, 2014 in prep.). Here, we will speak broadly 
about two classes of low-mass stars: partially convective and fully convective,
where the former are defined to have a convective envelope and radiative core.
Only radius inflation will be discussed, but effective temperatures have been
considered when reliable measurements are available.

\begin{figure}
	\centering
    \includegraphics[scale=0.30]{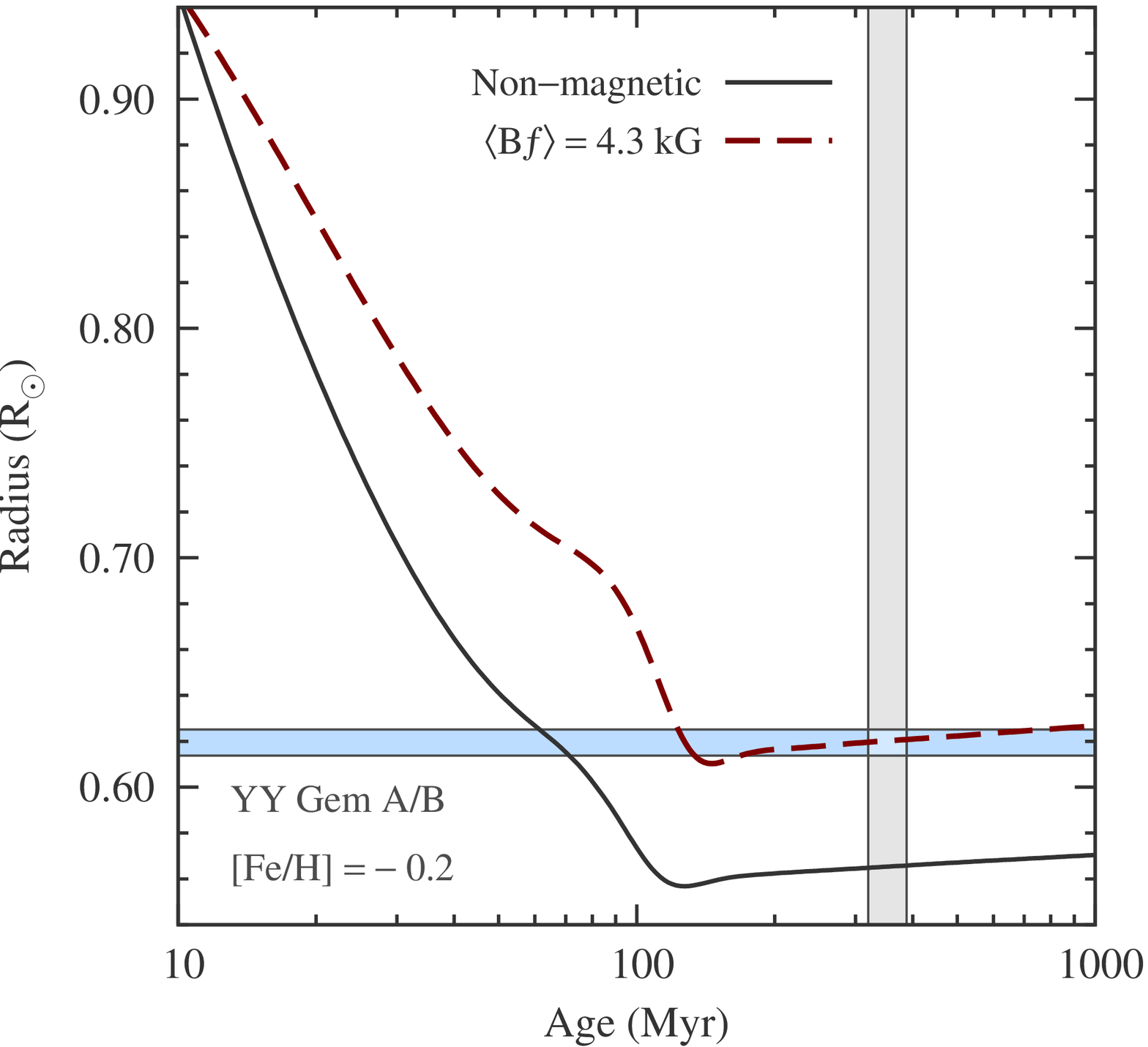}
    \qquad
    \includegraphics[scale=0.30]{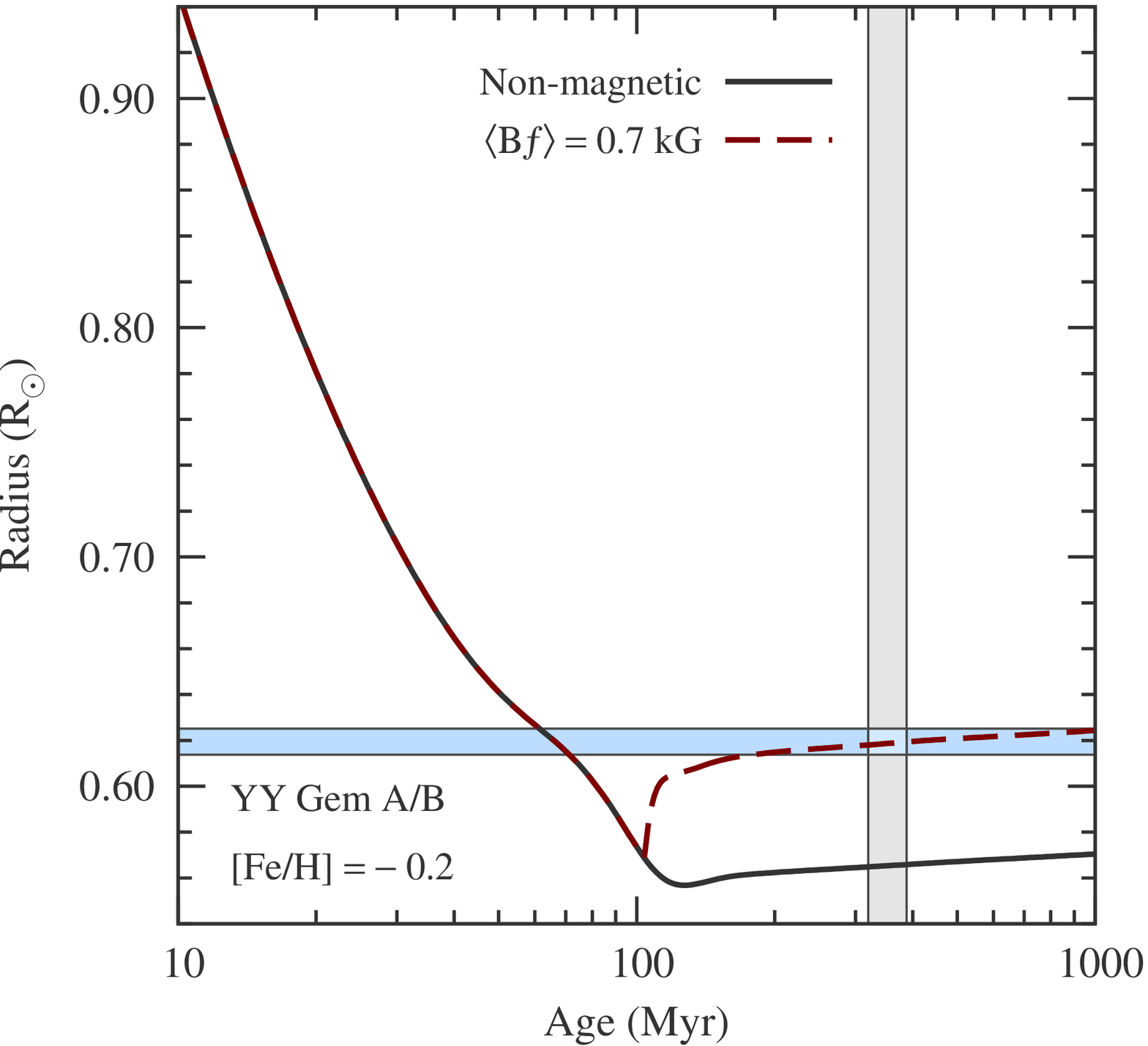}
    \caption{Radius evolution of standard (solid line) and magnetic 
             (dashed line) models of the equal-mass DEB YY Gem. 
             ({\it Left}) Magnetic
             model accounting for the stabilization of convection. ({\it Right})
             Magnetic model invoking inhibition of convective efficiency.
             The horizontal shaded region indicates the observed 
             radius with associated $1\sigma$ uncertainties and the 
             vertical region marks the estimated age.}
    \label{fig:yy_gem}
\end{figure}

An example of magneto-convection's impact on model radius predictions
for partially convective stars is given in Figure\,\ref{fig:yy_gem}. It
is quite clear from both panels that magneto-convection can provide a
proper solution, at least qualitatively. Model radii are forced to grow 
larger in the presence of a magnetic field. Notice, however, that the
two magneto-convection techniques yield different surface magnetic field 
strengths. In the left panel, we see that stabilization of convection
requires a surface magnetic field strength of 4.3\,kG whereas inhibition
of convective efficiency (right panel) only requires a 0.7\,kG surface 
magnetic field strength. These two values can be compared to surface magnetic
field strength estimates derived from the observed X-ray luminosity. 
In general, magneto-convection manifested as inhibition of convective efficiency
produces field strengths in better agreement with X-ray luminosity estimates
(Feiden \& Chaboyer \cite{FC13a}). It is therefore conceivable that magnetic fields
are inflating partially convective stars.

\begin{figure}
    \centering
    \includegraphics[scale=0.30]{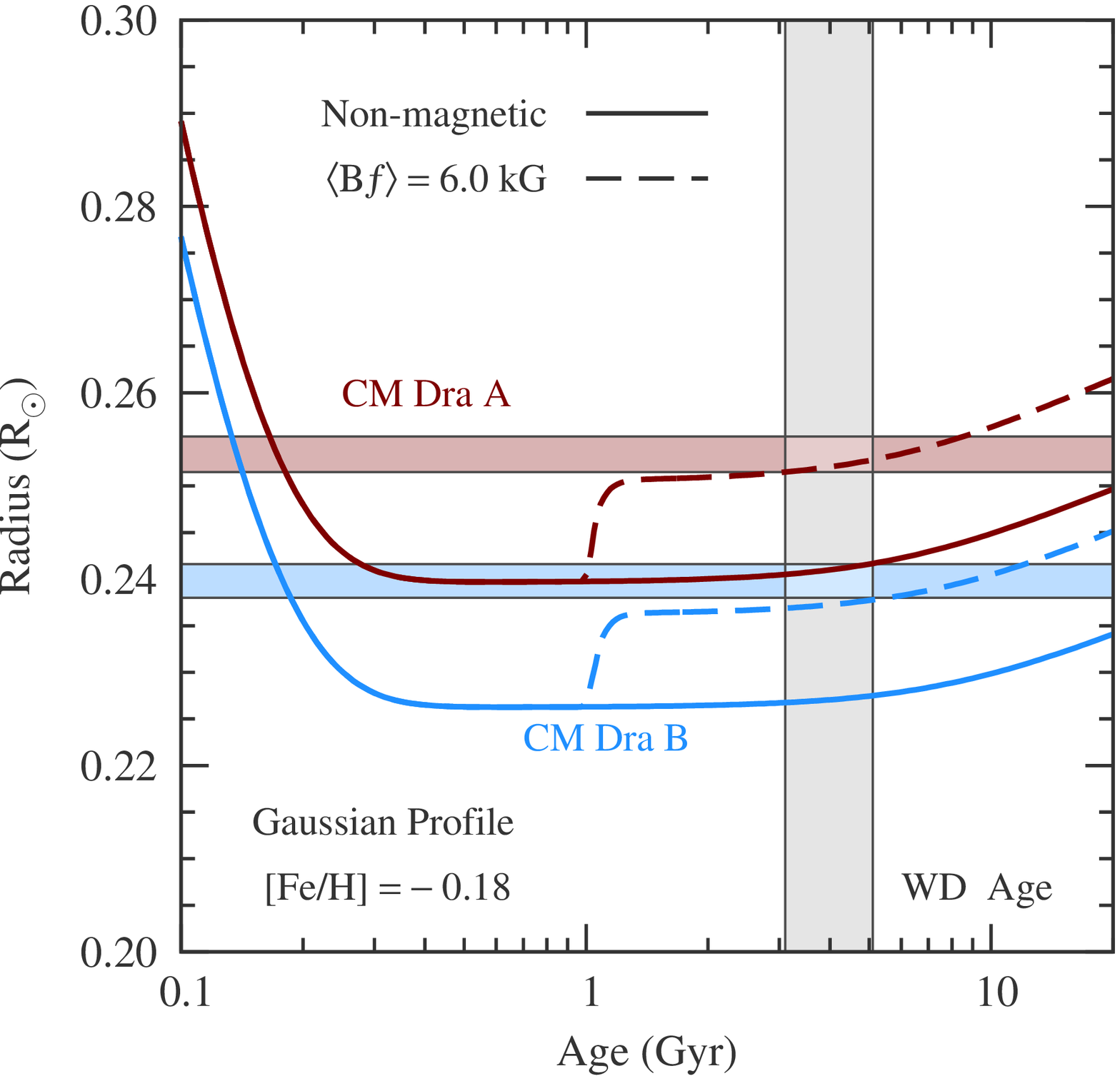}
    \qquad
    \includegraphics[scale=0.30]{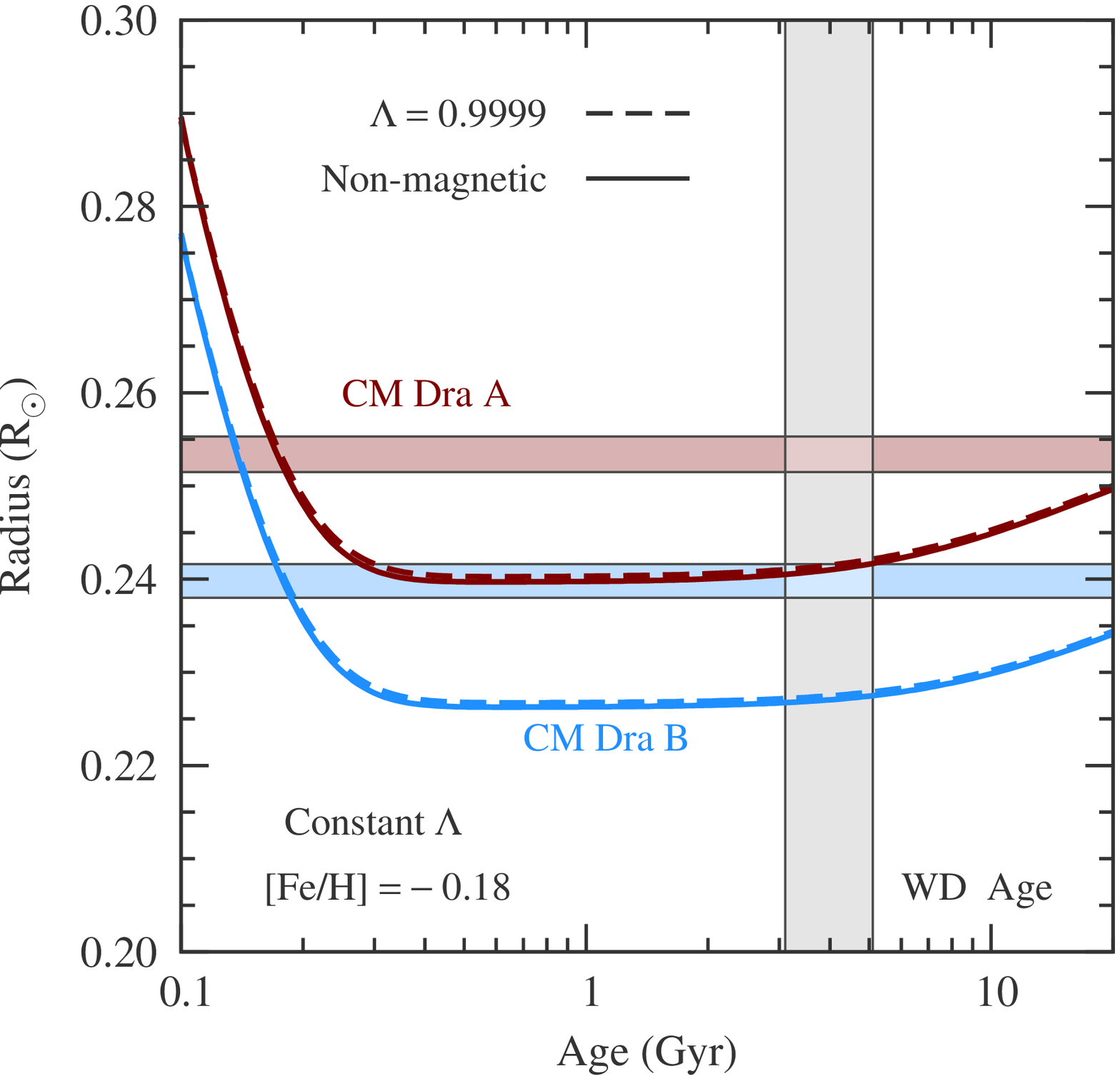}
    \caption{Same as Figure \ref{fig:yy_gem} for the fully convective 
             stars of CM Dra. (\textit{Left}) Models invoking the stabilization 
             of convection. (\textit{Right}) Models that have inhibited
             convective efficiency.}
    \label{fig:cm_dra}
\end{figure}

Fully convective stars appear more stubborn in the presence of a magnetic
field. Figure \ref{fig:cm_dra} shows that a qualitatively correct solution
can only be found when using stabilization of convection and a strong, 
nearly 6.0\,kG surface magnetic field strength. Inhibition of convective
inefficiency, on the other hand, has only a small effect on model radius 
predictions, even when assuming that 99.99\% of the kinetic energy 
in convection is converted to magnetic energy. It is important to note that
the surface magnetic field strengths used in the two methods are only
different by a factor of two (6.0\,kG versus 3.0\,kG), but the interior
magnetic field strengths differ by nearly three orders of magnitude (50\,MG
versus 50\,kG, respectively)! We have strong doubts about the existence of 50\,MG interior
magnetic fields. A 50\,kG magnetic field is easily obtained when
the magnetic field comes into equipartition with convective flows. However, we
have shown that inhibition of convective efficiency is unable to produce
agreement in the mass-radius plane. Additional discussion concerning magnetically
induced radius inflation in fully convective stars is presented in an upcoming
publication (Feiden \& Chaboyer 2014, in prep.), where we argue against the idea 
that magnetic fields are inflating fully convective stars.

\begin{figure}
    \centering
    \includegraphics[scale=0.30]{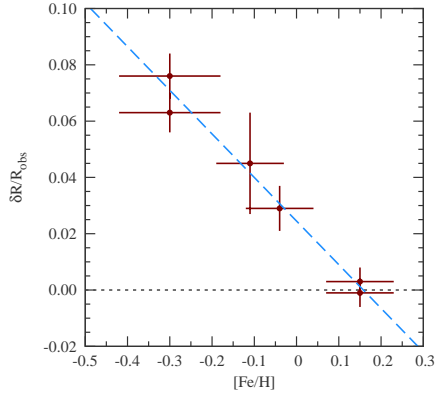}
    \caption{Relative radius error between stellar evolution models
             and observations of fully convective stars in DEBs plotted against
             observed stellar metallicities.}
    \label{fig:feh_resid}
\end{figure}

\section{Discussion}
Instead of magnetic fields inflating fully convective stars, we suggest that
observed discrepancies are related to metallicity, as is shown in Figure
\ref{fig:feh_resid}. Confirmation of this trend is required, which 
will require considerable effort to accurately determine DEB metallicities.
Some relief may come from systems where the primary star is an F, G, 
or K-dwarf. Additionally, direct measurements of surface magnetic field strengths
on well-studied DEBs would help confirm or falsify model predictions. Finally, 
reliable methods for deriving star spot properties are urgently needed. \\

{\it This work was supported by NSF grant AST-0908345 and the William
H.\ Neukom 1964 Institute for Computational Science at Dartmouth College.}


\end{document}